\documentclass[aoas,preprint]{imsart}
\pdfoutput=1

\RequirePackage{amsthm,amsmath,amsfonts,amssymb}
\RequirePackage[authoryear]{natbib}
\RequirePackage{graphicx}

\usepackage[utf8]{inputenc}
\usepackage[T1]{fontenc}
\usepackage{url}
\usepackage{booktabs}
\usepackage{amsfonts}
\usepackage{amsmath}
\usepackage{amsthm}
\usepackage{bm}
\usepackage{nicefrac}
\usepackage{microtype}
\usepackage{xcolor}
\usepackage{gnuplottex}
\usepackage{printlen}
\usepackage{enumerate}
\usepackage{hyperref}

\startlocaldefs
\newtheorem*{theorem}{Extremal Types Theorem}

\newcommand{\iid}{\mathbin{\overset{\text{\tiny{iid}}}{\sim}}}

\endlocaldefs

\makeatletter\begin{document}

\begin{frontmatter}
\title{Does Terrorism Trigger Online Hate Speech?\\ On the Association of Events and Time Series}
\runtitle{On the Association of Events and Time Series}

\begin{aug}
\author[]{\fnms{Erik} \snm{Scharw\"achter}\textsuperscript{*}\ead[label=e1]{scharwaechter@bit.uni-bonn.de}}
\and
\author[]{\fnms{Emmanuel} \snm{M\"uller}\ead[label=e2]{mueller@bit.uni-bonn.de}}
\address[]{Data Science and Data Engineering, University of Bonn, Germany;\\
\textsuperscript{*}scharwaechter@bit.uni-bonn.de, mueller@bit.uni-bonn.de
}
\end{aug}

\begin{abstract}
Hate speech is ubiquitous on the Web.
Recently, the offline causes that contribute to online hate speech have received increasing
attention.
A recurring question is whether the occurrence of extreme events offline
systematically triggers bursts of hate speech online, indicated by peaks in the volume of hateful
social media posts.
Formally, this question translates into measuring the association between
a sparse event series and a time series.
We propose a novel statistical methodology to measure, test and visualize the systematic association
between rare events and peaks in a time series.
In contrast to previous methods for causal inference or independence tests on time series,
our approach focuses only on the \textit{timing} of events and peaks, and no other distributional
characteristics.
We follow the framework of event coincidence analysis (ECA) that was originally developed to
correlate point processes.
We formulate a discrete-time variant of ECA and derive all required distributions to enable
analyses of peaks in time series, with a special focus on serial dependencies and peaks
over multiple thresholds.
The analysis gives rise to a novel visualization of the association via quantile-trigger rate plots.
We demonstrate the utility of our approach by analyzing whether Islamist terrorist attacks
in Western Europe and North America systematically trigger bursts of hate speech and counter-hate
speech on Twitter.
\end{abstract}


\end{frontmatter}

\section{Introduction}
\label{secIntro}

The ubiquity of hate speech in online social media, while distressing, delivers insights
into key emotive subjects within a society and globally.
The terms and conditions of most online social media platforms prohibit hate speech.
Providers ask users to report such contents in order to take further action,
e.g., by removing the contents, warning the involved users, or suspending or deleting their
profiles \citep{Matias2015}.
Deletion of harassing material and incitements to violence against individuals
is important to protect the victims. However, bursts of group-based hate speech online,
e.g., anti-Muslim, anti-immigrant, anti-black, antisemitic, or homophobic,
also help identifying triggers and mechanisms of hate and thereby inform policymakers and
non-governmental organizations.
A recent publication by the United Nations Educational, Scientific and Cultural Organization
(UNESCO) points out that the
``character of hate speech online and its relation to offline speech and action are poorly
understood'' and that the ``causes underlying the phenomenon and the dynamics through which certain
types of content emerge, diffuse and lead---or not---to actual discrimination, hostility or
violence'' should be investigated more deeply \citep{Gagliardone2015}.

We propose a novel statistical methodology that enables analyses of the systematic
relation between rare offline events and online social media usage.
Following a recent study \citep{Olteanu2018}, we demonstrate the utility of our approach by
analyzing whether Islamist terrorist attacks systematically trigger bursts of hate speech
and counter-hate speech on Twitter.
We operationalize these speech acts by tracking usage of the hashtags \texttt{\#stopislam}
(anti-Muslim hate speech) and \texttt{\#notinmyname} (Muslim counter-hate speech), as well as
the Arabic keyword \texttt{kafir} (jihadist hate speech against ``non-believers'') over a period
of three years (2015--2017).
We correlate usage of these terms with the occurrence of severe terrorist incidents in
Western Europe and North America in the same time period.
The key novelty of our approach is that we focus only on the
\textit{timing} of spikes in the resulting time series, not their \textit{magnitude} or
\textit{duration} as in previous studies \citep{Olteanu2018,Burnap2014}.
If spikes in the time series coincide with events more often than expected under an independence
assumption, there is evidence for a systematic statistical relationship between the two.
We thus map the correlation problem into the framework of event coincidence analysis (ECA)
\citep{Donges2016} that was recently proposed to measure coincidences for pairs of
point processes.

We first provide a discrete-time formulation of ECA for pairs of event series
that corresponds to the original continuous-time formulation for point processes.
Building on this formulation, our methodological contributions are as follows.
We replace the lagging event series with a thresholded time series and \textit{derive the null
distribution of the ECA statistic} for exceedances of a single threshold. The derivation
is valid for a large class of strictly stationary time series with serial dependencies.
Since a single threshold is often not sufficient to capture the association, we further
\textit{derive the joint distribution of the ECA statistic at multiple thresholds}.
The derivations yield two hypothesis tests for the association at multiple thresholds.
We further propose a novel visualization of trigger coincidences via
\textit{quantile-trigger rate plots} (QTR plots).
With our method, we are able to show that Islamist terrorist attacks in Western Europe
and North America systematically trigger bursts of anti-Muslim hate speech on Twitter
(\texttt{\#stopislam}), which confirms case studies and explorative analyses from the literature.
All source codes required to reproduce our results are available online at
\url{https://github.com/diozaka/pECA}.

\section{Related work}

Most research on the relation between offline actions/events and online hate speech so far is
based on \textit{case studies}.
The UNESCO publication mentioned earlier describes a few qualitative case studies
on the extreme right-wing online forum ``Stormfront''
\citep{DeKoster2008,Bowman-Grieve2009,Meddaugh2009}, and presents findings from non-academic
reports on online hate speech during elections in Kenya and against the Rohingya community in
Myanmar \citep{Gagliardone2015}.
\citet{Burnap2014} perform a quantitative case study of the social media reaction after the
Woolwich terrorist attack in the United Kingdom (May 23rd, 2013). They analyze the size and
survival rate of posts that express \textit{tension}, defined as antagonistic or accusatory
content similar to hate speech. In follow-up works \citep{Burnap2014a,Williams2016,Burnap2016},
they exploit their findings to train hate speech classifiers and predictive models for
information flow following emotive offline events.
\citet{Magdy2015} perform a quantitative case study of Twitter usage after
the Paris terrorist attacks (November 13th, 2015) and find 21.5\% of the posts attacking
Islam or Muslims, as opposed to 55.6\% defending posts. In contrast to these case studies,
we address the \textit{systematic} relation between offline events and online hate
speech in a longitudinal study with 17 relevant events over three years to uncover a potential
causal link.

\citet{Muller2019} empirically analyze the relation between online hate speech
and hate crimes against refugees in Germany by performing fixed effects panel regression on
data that covers the years 2015 to 2017.
They exploit internet outages as sources of quasi-experimental exogenous
variation to establish a causal link. The major difference to our work is that their events
of interest are so numerous that they can be aggregated to a numerical value with weekly
resolution and be analyzed with standard statistical methodology. In our setting, events are
very rare with respect to the daily resolution of the time series.

Most related to our work, \citet{Olteanu2018} analyze the impact of Islamist and Islamophobic
terrorist attacks on anti-Muslim hate speech online in a longitudinal study with 13 relevant
events over 19 months.
They perform counterfactual analyses \citep{Brodersen2015} of a large number of time series
representing the daily volumes of hundreds of anti-Muslim keywords independently for every
event and report aggregated effects.
However, the counterfactual approach is designed for singular, controlled interventions, not for
observational studies with reoccurring, uncontrolled events.
The reported aggregated effects are thus explorative and not corroborated by measures of
statistical significance. We fill this gap by providing a novel statistical methodology
to systematically analyze coincidences of rare events and peaks in a time series within the
framework of ECA.

ECA was developed to measure the association between two types of reoccurring events.
It was applied to assess whether floods systematically trigger epidemic
outbreaks \citep{Donges2016}, or whether natural disasters systematically trigger
violent conflicts \citep{Schleussner2016}. We give an introduction to ECA and discuss
challenges when applied to the study of peaks in time series in Section~\ref{secMethods}.
Event synchronization \citep{Quiroga2002} is similar to ECA, but allows
the time tolerance for coincidences to vary. This increases model expressiveness at the
cost of an analytical null distribution.
The major difference between our ECA-based approach and other methods for causal inference
in time series \citep{Granger1969,Box1975,Schreiber2000,Bressler2011,Brodersen2015} or
related independence tests \citep{Besserve2013,Chwialkowski2014,Scharwachter2020}
is that the only feature it uses is the \textit{timing} of events and peaks,
irrespective of other distributional characteristics.
In particular, it does not assume an underlying predictive model that would have
to explain the exact behavior of the time series after event occurrences.
By focusing on peaks in the time series, it is closely related to measures and models for tail
dependence of random variables \citep{Frahm2005,Yan2019a}.

Causal inference techniques have been applied in social media studies before
\citep{Cunha2017,Chandrasekharan2017,Saha2019}.
These works differ from ours in that they do not focus on the association between time series
and event series.
Recent work on online hate speech has focused on the targets of hate
\citep{Silva2016,Mondal2017,ElSherief2018}, characterizations of hateful users
\citep{ElSherief2018,Ribeiro2018}, as well as geographic \citep{Mondal2017} and linguistic
differences \citep{ElSherief2018a} in hate.
Perhaps the largest body of research on online hate speech in the past decade has been
on different approaches for its automatic identification
\citep{Warner2012,Kwok2013,Burnap2014a,Djuric2015,Davidson2017}.
An overview of the various approaches is given in a recent survey \citep{Schmidt2017}.

 \section{Data}
\label{secData}
For a quantitative analysis of social media usage in reaction to Islamist terrorist attacks
we have to operationalize these terms.

\subsection{Islamist terrorist attacks}
We obtained a comprehensive list of global terrorist attacks from the publicly available
Global Terrorism Database (GTD) \citep{START2018}.
We filtered the GTD for attacks that occurred in
Western Europe and North America between January 2015 and December 2017,
left at least 10~people wounded, and were conducted by the so-called \textit{Islamic State of Iraq
and the Levant} (ISIL), \textit{Al-Qaida in the Arabian Peninsula} (AQAP), Jihadi-inspired
extremists or Muslim extremists, according to the GTD.
The resulting 17~severe Islamist terrorist attacks are shown in Table~\ref{tblTerroristAttacks}.
\begin{table}
\caption{Severe Islamist terrorist attacks in Western Europe and North America.}
\label{tblTerroristAttacks}
\footnotesize
\begin{tabular}{l l c l l}
\toprule[1pt]
\textbf{Date} & \textbf{City} && \textbf{Date} & \textbf{City} \\
\cmidrule{1-2}
\cmidrule{4-5}
2015-01-07 & Paris, France  &&      2016-12-19 & Berlin, Germany \\
2015-11-13 & Paris, France  &&      2017-03-22 & London, UK \\
2015-12-02 & San Bernardino, USA && 2017-04-07 & Stockholm, Sweden \\
2016-03-22 & Brussels, Belgium &&   2017-05-22 & Manchester, UK \\
2016-06-12 & Orlando, USA  &&       2017-06-03 & London, UK \\
2016-07-14 & Nice, France  &&       2017-08-17 & Barcelona, Spain \\
2016-07-24 & Ansbach, Germany  &&   2017-09-15 & London, UK \\
2016-09-17 & New York City, USA &&  2017-10-31 & New York City, USA \\
2016-11-28 & Columbus, USA && \\
\bottomrule[1pt]
\end{tabular}\end{table}

\subsection{Social media response}
\begin{figure*}[tbp]
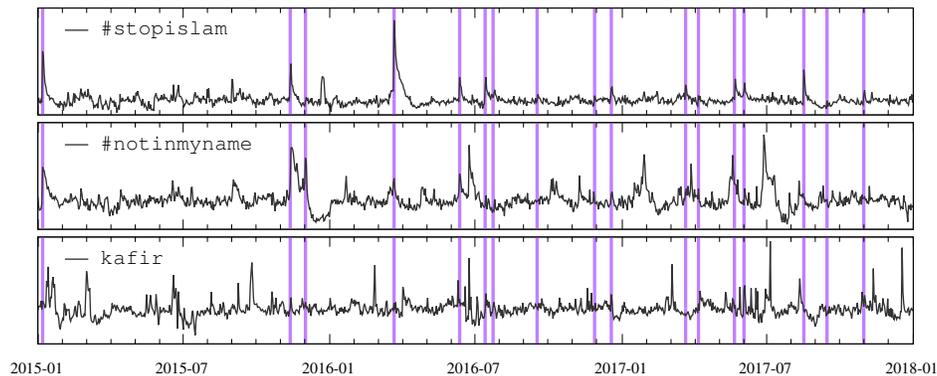

\begin{center}
\begin{gnuplot}[terminal=cairolatex,terminaloptions={size 4.9823,2.0}]
set multiplot layout 3,1
unset ytics
set xdata time
set timefmt "set xrange ["2015-01-01":"2018-01-01"]
set key left Left reverse samplen 1
#set grid xtics
#set grid mxtics
set grid front # brings xtics to front
unset grid

set tmargin at screen 0.98
set bmargin at screen 0.70
set format x ""
set yrange [-2:12]
plot "data/events-wena-min10wound.dat" u 1:($2*-1000) w i lc 'purple' lw 5 notitle, \
     "data/events-wena-min10wound.dat" u 1:($2*1000) w i lc 'purple' lw 5 notitle, \
     "data/hatespeech-daily-transformed.dat" u 1:2 w l lc rgb '#333333' lw 2 title '\scriptsize{\texttt{\#stopislam}}'

set tmargin at screen 0.68
set bmargin at screen 0.40
set yrange [-3.5:9]
plot "data/events-wena-min10wound.dat" u 1:($2*-1000) w i lc 'purple' lw 5 notitle, \
     "data/events-wena-min10wound.dat" u 1:($2*1000) w i lc 'purple' lw 5 notitle, \
     "data/hatespeech-daily-transformed.dat" u 1:3 w l lc rgb '#333333'  lw 2 title '\scriptsize{\texttt{\#notinmyname}}'

set tmargin at screen 0.38
set bmargin at screen 0.10
set format x '\tiny{set yrange [-1:2]
plot "data/events-wena-min10wound.dat" u 1:($2*-1000) w i lc 'purple' lw 5 notitle, \
     "data/events-wena-min10wound.dat" u 1:($2*1000) w i lc 'purple' lw 5 notitle, \
     "data/hatespeech-daily-transformed.dat" u 1:4 w l lc rgb '#333333'  lw 2 title '\scriptsize{\texttt{kafir}}'

\end{gnuplot}
\caption{Daily Twitter volume of the keywords analyzed in this study. The vertical
lines indicate dates of severe Islamist terrorist attacks in Western Europe and North America.}
\label{figData}
\end{center}
\end{figure*}

To assess the online social media response to these events in terms of anti-Muslim hate speech,
Muslim counter-hate speech, and jihadist hate speech, we retrieved time series of the
global Twitter volume in the same time period (2015--2017) for the three keywords
\texttt{\#stopislam}, \texttt{\#notinmyname} and \texttt{kafir} (``non-believer'') that represent
the three speech acts.
We retrieved daily time series of the global Twitter volume for our keywords.
Details on keyword selection, data acquisition and preprocessing are given in
Appendix~\ref{secDetData}.
The global daily volume for all queries after preprocessing is shown in Figure~\ref{figData},
along with all Islamist terrorist attacks from Table~\ref{tblTerroristAttacks}.

 \section{Methods}
\label{secMethods}

\begin{figure}[tbp]
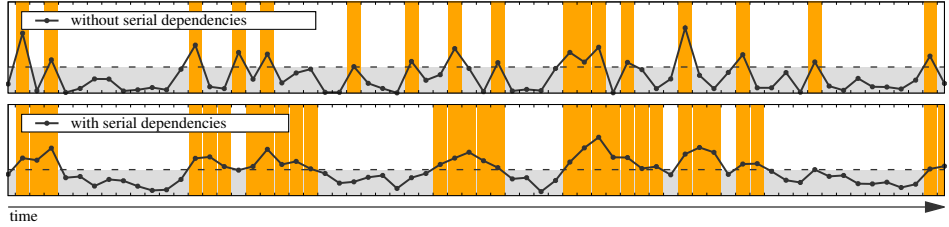

\begin{center}
\begin{gnuplot}[terminal=cairolatex,terminaloptions={size 4.9823,1.2 dashed}]
set multiplot layout 2,1
set format x ''
unset xtics
set xtics -0.5,1 scale 0.3
unset ytics
set lmargin 0.01
set rmargin 0.99
set xrange [5:70]
set yrange [0:7]
set style fill solid noborder
set boxwidth 0.9

set grid front # brings xtics to front
unset grid

set key top left reverse Left samplen 1 box opaque width -10 height -0.1 spacing 0.8
set arrow from 5, screen 0.05 to 70, screen 0.05 lw 2 lc rgb '#333333' head filled
set label '\tiny{time}' at 5.2, screen 0.01
set tmargin screen 0.95
set bmargin screen 0.55
#set label '\tiny{without serial dependencies}' at 10, 4.7 front
plot 2.0 w filledcurves x1 dt 2 lw 2 lc rgb '#DDDDDD' notitle, \
     "data/exponential.dat" u (($1>2.0)*10) w boxes notitle fc 'orange' lw 3, \
     "data/exponential.dat" w lp lc '#333333' lw 3 pt 7 ps 0.2 title '\tiny{without serial dependencies}', \
     2.0 dt 2 lw 2 lc rgb '#333333' notitle

set key top left reverse Left samplen 1 box opaque width -9 height -0.1 spacing 0.8
unset arrow
unset label
set tmargin screen 0.50
set bmargin screen 0.10
#set label '\tiny{with serial dependencies}' at 10, 4.7 front
plot 2.0 w filledcurves x1 dt 2 lw 2 lc rgb '#DDDDDD' notitle, \
     "data/exponential-roll.dat" u (($1>2.0)*10) w boxes notitle fc 'orange' lw 3, \
     "data/exponential-roll.dat" w lp lc '#333333' lw 3 pt 7 ps 0.2 title '\tiny{with serial dependencies}', \
     2.0 dt 2 lw 2 lc rgb '#333333' notitle
\end{gnuplot}
\end{center}
\caption{Threshold exceedance series (bars) for time series with and
without serial dependencies (lines).}
\label{figThreshX}
\end{figure}

Our goal is to analyze the systematic relation between offline events and online social media usage.
Formally, we model the occurrence of terrorist attacks by a (discrete-time) \textbf{event series}
$\mathcal{E} = (E_t)_{t=1}^{T}$, where each $E_t$ is a binary random variable with $E_t = 1$
if and only if there is a terrorist attack at time $t$, and $E_t=0$ otherwise.
Social media usage is captured by a (discrete-time) \textbf{time series}
$\mathcal{X} = (X_t)_{t=1}^{T}$, where each $X_t$ is a continuous random variable that
indicates the daily volume of posts.
A \textbf{peak} in the time series is the exceedance of some large threshold $\tau \in \mathbb{R}$.
The problem is to decide whether the number of events in $\mathcal{E}$ that
trigger peaks in $\mathcal{X}$ is so high that the association should be considered statistically
significant: in this case, there is a potential causal link between event occurrences and peaks
in the time series.
Observe that for a time series $\mathcal{X}$ and threshold $\tau$, the
\textbf{threshold exceedance series}
\begin{equation}
\label{eqnExceedanceProcess}
\mathcal{A} = \left(\mathcal{I}(X_1 > \tau), ..., \mathcal{I}(X_T > \tau)\right)
\end{equation}
is itself an event series. Here, $\mathcal{I}(C)$ is an indicator function that is 1
if and only if the condition $C$ is true, and 0 otherwise. See~Figure~\ref{figThreshX}
for an example.
The threshold exceedance series retains only information on the timing of exceedances,
and disregards all other distributional characteristics.
The problem of correlating event series with peaks in a time series can thus directly be mapped
to the problem of correlating two event series, e.g., using ECA.

There are two challenges that need to be addressed when applying measures designed for
pairs of event series in this context:
\textbf{serial dependencies} and \textbf{threshold selection}.
Serial dependencies in the time series lead to clustering of events in the threshold exceedance
series. This effect can be observed when comparing the upper and the lower time series in
Figure~\ref{figThreshX}.
Event clustering must be handled correctly when establishing the statistical significance
of an observed correlation score.
The second challenge is that the choice of threshold has a strong impact on the results of the
analysis, but is often not straight-forward. In fact, the magnitude of the response may vary from
event to event: a full picture of the association between events and peaks can only be obtained
when considering exceedances at multiple thresholds.

We now proceed with a detailed exposition of the statistical methodology that we propose
for the analysis. We embed our contributions within the existing framework of ECA.
We begin with a discrete-time formulation of ECA for pairs of event series in Section~\ref{secDECA}
that corresponds to the original continuous-time formulation for point processes \citep{Donges2016}.
In Section~\ref{secTriggerDistrFixedThresh} we address the challenge of serial dependencies
by deriving a novel analytical null distribution for the ECA statistic that is valid for threshold
exceedance series for a large class of strictly stationary time series.
We then derive the joint null distribution for coincidences at multiple thresholds in
Section~\ref{secIncreasingThresholds} and describe two test procedures to assess statistical
significance. We complement the analytical results with a novel visualization of the
association via quantile-trigger rate (QTR) plots.

\subsection{Discrete-time event coincidence analysis}
\label{secDECA}

ECA is a statistical methodology to assess whether two types of events are independent or
whether one kind of event systematically triggers or precedes the other kind of event.
The basic idea of ECA is to count how many times the two kinds of events coincide, and assess
whether this number is statistically significant under the assumption of independence.

\subsubsection{Definition}
\begin{figure}[tbp]
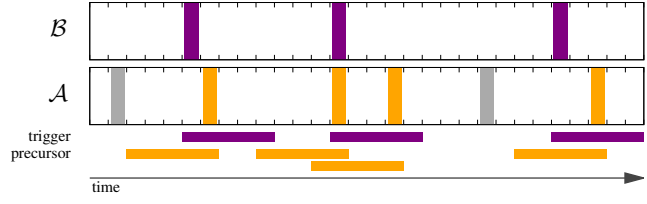

\begin{center}
\begin{gnuplot}[terminal=cairolatex,terminaloptions={size 3.3374,1.0}]
set multiplot layout 2,1
set style fill solid noborder
set boxwidth 0.75
unset ytics
set xtics -0.5,1 scale 0.5
set xrange [0.5:30.5]
set format x ''
#set grid xtics
set lmargin 6
set rmargin 0.1

set obj rect from  5.5, screen 0.27 to 10.5, screen 0.32 fc '#800080' fs solid 1 noborder
set obj rect from 13.5, screen 0.27 to 18.5, screen 0.32 fc '#800080' fs solid 1 noborder
set obj rect from 25.5, screen 0.27 to 30.5, screen 0.32 fc '#800080' fs solid 1 noborder
set obj rect from  2.5, screen 0.18 to  7.5, screen 0.23 fc 'orange' fs solid 1 noborder
set obj rect from  9.5, screen 0.18 to 14.5, screen 0.23 fc 'orange' fs solid 1 noborder
set obj rect from 12.5, screen 0.12 to 17.5, screen 0.17 fc 'orange' fs solid 1 noborder
set obj rect from 23.5, screen 0.18 to 28.5, screen 0.23 fc 'orange' fs solid 1 noborder
set label '\tiny{trigger}' right at 0.5, screen 0.30 offset -1.3
set label '\tiny{precursor}' right at 0.5, screen 0.21 offset -1.3

set arrow from 0.5, screen 0.08 to 30.5, screen 0.08 lw 2 lc rgb '#444444' head filled
set label '\tiny{time}' at 0.7, screen 0.03

set tmargin at screen 1
set bmargin at screen 0.7
set ylabel '$\mathcal{B}$ ' rotate by 0
plot "data/eca-toy.dat" u 1 w boxes notitle fc rgb '#AAAAAA', \
     "data/eca-toy.dat" u 3 w boxes notitle fc '#800080'

unset obj
unset label
unset arrow
set tmargin at screen 0.66
set bmargin at screen 0.36
set ylabel '$\mathcal{A}$ '
plot "data/eca-toy.dat" u 2 w boxes notitle fc rgb '#AAAAAA', \
     "data/eca-toy.dat" u 4 w boxes notitle fc 'orange'
\end{gnuplot}
\end{center}
\caption{Trigger coincidences and precursor coincidences for two event series
$\mathcal{A}$ and $\mathcal{B}$, with time tolerance $\Delta=4$.}
\label{figDECACoincidences}
\end{figure}

Let $\mathcal{A} = (E_t^\mathcal{A})_{t=1}^{T}$ and $\mathcal{B} = (E_t^\mathcal{B})_{t=1}^{T}$
be two event series of length $T$ with $E^\mathcal{A}_t, E^\mathcal{B}_t \in \{0,1\}$ for all $t$,
and $N_\mathcal{A} = \sum_t E_t^\mathcal{A}$ and $N_\mathcal{B} = \sum_t E_t^\mathcal{B}$
event occurrences. Furthermore, let $\Delta \in \mathbb{N}_0$ be a user-defined
time tolerance. ECA measures the extent to which $B$ events precede $A$ events,
with a time tolerance of $\Delta$.
We thus refer to $\mathcal{A}$ as the \textit{lagging} and $\mathcal{B}$ the \textit{leading}
event series.
ECA considers two possibilities to measure this extent:
\textbf{trigger coincidences} and \textbf{precursor coincidences}.
A trigger coincidence occurs whenever a $\mathcal{B}$ event triggers an $\mathcal{A}$ event
within the next $\Delta$ time steps, whereas a precursor coincidence occurs whenever an
$\mathcal{A}$ event is preceded by a $\mathcal{B}$ event within the previous $\Delta$ time steps.
In the special case $\Delta=0$, the two concepts are identical.
The two types of coincidences are illustrated in Figure~\ref{figDECACoincidences}
with $\Delta=4$.
In the example, there are three trigger coincidences and four precursor coincidences.
A significant number of trigger or precursor coincidences indicates a possible causal link
from $\mathcal{B}$ to $\mathcal{A}$.
The opposite direction can be analyzed analogously by exchanging the labels.
Formally, the \textbf{number of trigger coincidences} is defined as
\begin{gather}
\label{eqnDECATriggerCoincidences}
K_\text{tr} = K^{\Delta}_{\text{tr}}(\mathcal{B},\mathcal{A}) := \sum_{t=1}^{T-\Delta} E^\mathcal{B}_t \cdot \left(\max_{\delta=0,...,\Delta} E^\mathcal{A}_{t+\delta}\right)
\end{gather}
and the \textbf{number of precursor coincidences} as
\begin{gather}
\label{eqnDECAPrecursorCoincidences}
K_\text{pre} = K^{\Delta}_{\text{pre}}(\mathcal{B},\mathcal{A}) := \sum_{t=\Delta+1}^{T} E^\mathcal{A}_t \cdot \left(\max_{\delta=0,...,\Delta} E^\mathcal{B}_{t-\delta}\right).
\end{gather}
The order of the function arguments $\mathcal{B}$ and $\mathcal{A}$ corresponds to the temporal
ordering that is analyzed (and thus the potential causal direction).
We omit the parameter $\Delta$ and the function arguments whenever they are clear from the context.
The corresponding \textbf{coincidence rates} are given by
$r_\text{tr} := K_\text{tr}/N_\mathcal{B}$ and
$r_\text{pre} := K_\text{pre}/N_\mathcal{A}$.
In the example from Figure~\ref{figDECACoincidences}, we have
$r_\text{tr} = 1$ and $r_\text{pre} = \frac{2}{3}$.
A high \textit{trigger coincidence rate} indicates that a large fraction of $\mathcal{B}$ events
is followed by an $\mathcal{A}$ event. In other words, $\mathcal{B}$ events systematically
trigger $\mathcal{A}$ events.
A high \textit{precursor coincidence rate} indicates that a large fraction of $\mathcal{A}$ events
is preceded by a $\mathcal{B}$ event, i.e., the occurrence of $\mathcal{A}$ events can be explained
to a large degree by $\mathcal{B}$ events.
The two measures are complementary and should be selected based on the research question.

\subsubsection{Null distribution}
\label{secDECANullDistr}
To assess whether an observed trigger coincidence rate is statistically
significant, we need the null distribution of $K_\text{tr}$ under the assumption that the processes
are independent.
For this purpose, we introduce the binary random variables
$Z^\mathcal{A}_t := \max_{\delta=0,...,\Delta} E^\mathcal{A}_{t+\delta}$ for all $t = 1,...,T-\Delta$
that indicate whether there is an $\mathcal{A}$ event in the window $t,...,t+\Delta$.
This allows rewriting Equation~\ref{eqnDECATriggerCoincidences} as
\begin{gather}
K_\text{tr} = \sum_{t=1}^{T-\Delta} E^\mathcal{B}_t \cdot Z^\mathcal{A}_t
= \sum_{t: E^\mathcal{B}_t=1} Z^\mathcal{A}_t
\end{gather}
and reveals that the number of trigger coincidences is effectively a sum of Bernoulli trials, each
associated with an event occurrence in $\mathcal{B}$.
Sums over fixed numbers of independent and identically distributed Bernoulli trials follow
binomial distributions.
However, for an arbitrary event series $\mathcal{A}$, the random variables
$Z^\mathcal{A}_t$ and $Z^\mathcal{A}_{t'}$ may be neither
identically distributed nor independent. Additional assumptions are required to derive the
null distribution analytically.

In a simple and analytically tractable case, the event series $\mathcal{A}$ and $\mathcal{B}$
are independent Bernoulli processes with
$\mathbb{P}(E^\mathcal{A}_t = 1) = p_\mathcal{A}$ and $\mathbb{P}(E^\mathcal{B}_t = 1)
   = p_\mathcal{B}$ for all $t$.
In this case, all $Z^\mathcal{A}_t$ are identically distributed with success probability
\begin{align}
\label{eqnDECAindicators}
\mathbb{P}(Z^\mathcal{A}_t = 1) = 1-\mathbb{P}(E^\mathcal{A}_t=0,...,E^\mathcal{A}_{t+\Delta}=0)
 = 1 - (1-p_\mathcal{A})^{\Delta+1}.
\end{align}
Furthermore, two variables $Z^\mathcal{A}_t$ and $Z^\mathcal{A}_{t'}$ are independent whenever
they are separated by more than $\Delta$ time steps.
Under the additional assumption that the $N_\mathcal{B}$ events in $\mathcal{B}$ are
separated by more than $\Delta$ time steps, the null distribution of
$K_\text{tr}$ is the binomial
\begin{equation}
\label{eqnDECANullDistrCond}
K_\text{tr} \mid N_\mathcal{B} \sim \operatorname{Binomial}\left(N_\mathcal{B}, 1 - (1-p_\mathcal{A})^{\Delta+1}\right).
\end{equation}

\subsubsection{Statistical test procedure}
Following the framework of ECA, we use $K_\text{tr}$ as a test statistic to decide
between the null hypothesis of independence of $\mathcal{A}$ and $\mathcal{B}$, and the
alternative hypothesis of a trigger relationship.
If the number of trigger coincidences is unusually large, the null hypothesis is rejected in
favor of the alternative hypothesis.
The success probabilities are estimated as $\hat{p}_\mathcal{A} = N_\mathcal{A}/T$ and
$\hat{p}_\mathcal{B} = N_\mathcal{B}/T$, respectively.
The $p$-value for an observed number of trigger coincidences $k_\text{tr}$ is obtained from
the probability mass function of the binomial distribution in Equation~\ref{eqnDECANullDistrCond},
\begin{equation}
\mathbb{P}(K_\text{tr} \ge k_\text{tr} \mid N_\mathcal{B})
= \sum_{k=k_\text{tr}}^{N_\mathcal{B}} \binom{N_\mathcal{B}}{k}
    \cdot \pi^k \cdot (1-\pi)^{N_\mathcal{B}-k},
\label{eqnDECApval}
\end{equation}
where $\pi = 1 - (1-\hat{p}_\mathcal{A})^{\Delta+1}$. The null hypothesis is rejected
at the desired significance level~$\alpha$
if $\mathbb{P}(K_\text{tr} \ge k_\text{tr} \mid N_\mathcal{B}) < \alpha$.
The null distribution and test for significance for the number of precursor coincidences can
be derived completely analogously.
Equation~\ref{eqnDECApval} is valid for event series that follow Bernoulli processes;
they correspond to the null distributions derived for homogeneous
Poisson processes in continuous-time ECA \citep{Donges2016}.
For other processes, such analytical results have not been obtained, and
to date, Monte Carlo methods are required to simulate the null distribution.

\subsection{Coincidences with threshold exceedances}
\label{secTriggerDistrFixedThresh}

Threshold exceedance series are typically not Bernoulli processes.
We now derive novel analytical results when the lagging event series is a
threshold exceedance series.
Our key observation is that $K_\text{tr}$ can be reformulated such that the
Extremal Types Theorem \citep{Coles2001} is applicable.
We first define the number of trigger coincidences for a leading event series $\mathcal{E}$,
lagging time series $\mathcal{X}$, time tolerance $\Delta \in \mathbb{N}_0$, and a threshold
$\tau \in \mathbb{R}$ by substituting the threshold exceedance series into
Equation~\ref{eqnDECATriggerCoincidences}:
\begin{equation}
\label{eqnTriggerCoincidences}
K_\text{tr} = K^{\Delta,\tau}_{\text{tr}}(\mathcal{E},\mathcal{X}) := \sum_{t=1}^{T-\Delta} E_t
      \cdot \left(\max_{\delta=0,...,\Delta} \mathcal{I}(X_{t+\delta}>\tau)\right).
\end{equation}
Observe that we can now swap the order of the max-operator and the indicator function
\begin{equation}
\max_{\delta=0,...,\Delta} \mathcal{I}(X_{t+\delta}>\tau)
= \mathcal{I}\left(\left(\max_{\delta=0,...,\Delta} X_{t+\delta}\right)>\tau\right)
\end{equation}
and introduce the helper variables
\begin{gather}
Z^{\Delta,\tau}_t :=
\mathcal{I}\left(\left(\max_{\delta=0,...,\Delta}X_{t+\delta}\right)  > \tau\right)
\end{gather}
such that
\begin{align}
K_{\text{tr}} = \sum_{t=1}^{T-\Delta} E_t \cdot Z^{\Delta,\tau}_t
= \sum_{t: E_t=1} Z^{\Delta,\tau}_t \label{eqnTriggerCoincidencesMaxSimple}.
\end{align}
The number of trigger coincidences is again a sum of Bernoulli trials, each associated with
one of the $N_\mathcal{E}$ event occurrences in $\mathcal{E}$.
If the time series $\mathcal{X}$ is strictly stationary, the Bernoulli trials are all identically
distributed with the same marginal distribution $\mathbb{P}(Z^{\Delta,\tau}_t)$, but they are,
in general, not independent.
The benefit of swapping the max-operator and the indicator function is that now the success
probability of the Bernoulli trials
\begin{align}
\mathbb{P}(Z^{\Delta,\tau}_t = 1)
 = \mathbb{P}\left(\max_{\delta=0,...,\Delta} X_{t+\delta} > \tau\right)
 = 1 - \mathbb{P}\left(\max_{\delta=0,...,\Delta} X_{t+\delta} \le \tau\right)
 \label{eqnTriggerCoincidencesIndicator}
\end{align}
is defined directly on the time series, not on the event series as in
Equation~\ref{eqnDECAindicators}. Probabilities in the form of
Equation~\ref{eqnTriggerCoincidencesIndicator} have been studied extensively in
Extreme Value Theory. In fact, according to the Extremal Types Theorem (ETT), the maximum of
a large number of iid random variables is approximated by the Generalized Extreme Value (GEV)
distribution, under mild constraints on the underlying distribution of the random variables:
\begin{theorem}[ETT, \cite{Coles2001}]
Let $X_1, ..., X_n \iid F$ and $M_n = \max_{i=1,...,n} X_i$. If there exist sequences
of constants $\{a_n > 0\}$ and $\{b_n\}$ such that
\[
\mathbb{P}\left(\frac{M_n-b_n}{a_n} \le z\right)
  \longrightarrow G(z) \text{ as } n \longrightarrow \infty
\]
for a non-degenerate distribution function $G$, then $G$ is a member of the GEV family
\[
  G(z) = \exp\left\{-\left[1 + \xi\left(\frac{z-\mu}{\sigma}\right)\right]^{-\frac{1}{\xi}}\right\},
\]
defined on $\{z : 1 + \xi(z-\mu)/\sigma > 0\}$, where $-\infty < \mu < \infty$, $\sigma > 0$
and $-\infty < \xi < \infty$.
\end{theorem}
The ETT was shown to also apply more generally to the maxima of strictly
stationary time series, if they fulfill a regularity condition that eliminates
long-range dependencies; see \citet[ch.~5.2]{Coles2001} for technical details.
Under the conditions of the ETT and for large $\Delta$,
the variables $Z^{\Delta,\tau}_t$ are thus identically distributed with
\begin{align}
\mathbb{P}(Z^{\Delta,\tau}_t=1)
    &{}\approx 1 - G(\tau ; \bm{\theta}_\Delta).
\label{eqnBlockExceedanceGEV}
\end{align}
The normalizing constants from the ETT disappear in the GEV parameter vector
$\bm{\theta}_\Delta = (\xi, \mu, \sigma)$ that depends on $\Delta$.
The parameters are estimated by splitting the time series $\mathcal{X}$ into consecutive blocks
of size $\Delta+1$ and fitting the GEV distribution to the maxima of each block, e.g.,
by maximum likelihood.
The larger $\Delta$, the better the approximation by the GEV
distribution.\footnote{In Section~\ref{secNullDistributionPerformance}, we demonstrate that
a value of $\Delta=7$ can be large enough for good approximations, if the threshold $\tau$
is a high quantile of the data.}

This allows us to state the key result of this section:
If the conditions of the ETT hold, and events in $\mathcal{E}$ are sparse
(such that the variables $Z^{\Delta,\tau}_t$ associated with the $N_\mathcal{E}$ event occurrences
in $\mathcal{E}$ are approximately independent),
then the null distribution of the number of trigger coincidences for large $\Delta$
is approximated by the binomial
\begin{equation}
\label{eqnKtrGEV}
K^{\tau,\Delta}_\text{tr} \mid N_\mathcal{E} \sim \operatorname{Binomial}(N_\mathcal{E},
    1 - G(\tau ; \bm{\theta}_\Delta)).
\end{equation}
The statistical significance of an observed number of trigger coincidences $k_\text{tr}$
can hence be assessed using the $p$-value from Equation~\ref{eqnDECApval} with
$\pi = 1 - G(\tau ; \bm{\theta}_\Delta)$.

\subsection{Examples}
With these results, we are able to apply discrete-time ECA to analyze the triggers
of exceedances of a \textit{single} threshold.
The threshold may be derived from domain-specific hypotheses.
For example, we might want to test whether Islamist terrorist attacks systematically trigger
bursts of more than 1,000 posts per day on Twitter that contain the hashtag \texttt{\#stopislam}.
We observe $K_\text{tr} = 9$ trigger coincidences within a time tolerance of
$\Delta=7$ days after the events, which gives a trigger coincidence rate of $r_\text{tr}= .53$.
Assuming that the raw Twitter time series fulfills the conditions of the ETT,
we obtain a $p$-value of $p = .0014$ using our GEV-based null distribution.
Lacking a specific hypothesis for the value of the threshold, generic values can be tested,
such as the empirical 95\%-quantile.
For example, we can test the hypothesis that Islamist terrorist attacks systematically trigger
burstsof \texttt{\#notinmyname} usage that exceed the volume of 95\% of all days.
We observe $K_\text{tr} = 4$ trigger coincidences with $\Delta=7$, which gives a
trigger coincidence rate of $r_\text{tr}= .24$.
Again assuming that the raw Twitter time series fulfills the ETT conditions,
we obtain a $p$-value of $p = .2427$.
These examples are only illustrative, since the raw Twitter time series are not
strictly stationary and thus do not fulfill the ETT conditions.
In Appendix~\ref{secDetData}, we describe the preprocessing scheme that
we use for the results in Section~\ref{secResults} to make the time series stationary.

\subsection{Exceedances of increasing thresholds}
\label{secIncreasingThresholds}

In case a reasonable threshold is unknown, or if a full picture of the association with
peaks of various magnitudes is required, exceedances at multiple thresholds have to be considered.
Threshold exceedances at multiple levels are highly
dependent: if an observation exceeds any threshold $\tau$, it also exceeds all lower thresholds.
The numbers of trigger coincidences at multiple thresholds are thus dependent as well.
To enable joint analyses of multiple threshold exceedances and thereby eliminate
the need of selecting a single fixed threshold, we now derive the joint null distribution
of trigger coincidences at multiple thresholds, and provide a novel visualization for this
statistical association.

\subsubsection{Trigger coincidence processes}
\label{secTCP}
\begin{figure}[tbp]
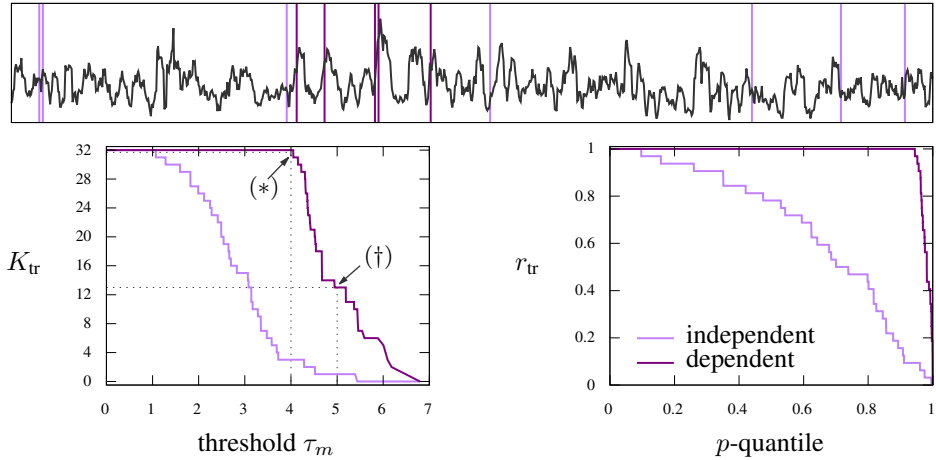

\begin{center}
\begin{gnuplot}[terminal=cairolatex,terminaloptions={size 4.9823,2.5 dashed}]
#set multiplot layout 2,1
set multiplot

set grid front # brings xtics to front
unset grid

unset key

set xrange [0:1024]

set lmargin screen 0.01
set rmargin screen 0.98
set tmargin screen 0.95
set bmargin screen 0.70
set yrange [0:7]
set format x ''
unset ytics
unset xtics
set format y '\tiny{ plot 'data/tcp-ex.dat' u 1:($3*100) w i lc 'purple' lw 3 notitle, \
     'data/tcp-ex.dat' u 1:($4*100) w i lc '#800080' lw 3 notitle, \
     'data/tcp-ex.dat' u 1:2 w l lc '#333333' lw 3 notitle

set tmargin screen 0.65
set bmargin screen 0.15

unset xrange
unset yrange
set xrange [0:7]
set xtics 0, 1
set format x '\tiny{set xlabel '\small{threshold $\tau_m$}'
set yrange [-0.5:32.5]
set ytics 0,4
set ylabel '\small{$K_\text{tr}$}' rotate by 0 offset 3
set format y '\tiny{set lmargin screen 0.11
set rmargin screen 0.45
set arrow from 3.5, 28 to 3.95, 31.0 lw 2 lc rgb '#333333' head filled front
set arrow from 4, 0 to 4, 32 lw 1 dt 3 lc rgb '#333333' nohead
set arrow from 0, 31.7 to 4, 31.7 lw 1 dt 3 lc rgb '#333333' nohead
set label '$(\ast)$' at 3.1, 26.5
set arrow from 5.5, 16 to 5.05, 13.5 lw 2 lc rgb '#333333' head filled front
set arrow from 5, 0 to 5, 13 lw 1 dt 3 lc rgb '#333333' nohead
set arrow from 0, 13 to 5, 13 lw 1 dt 3 lc rgb '#333333' nohead
set label '$(\dagger)$' at 5.6, 17
#plot 'data/tcp-ex-coinc.dat' u 1:4 w l lc '#333333' lw 6 dt 2 title 'expected'
plot 'data/tcp-ex-coinc.dat' u 1:2 w l lc 'purple'  lw 3 title 'independent', \
     'data/tcp-ex-coinc.dat' u 1:3 w l lc '#800080'    lw 3 title 'dependent'
unset label
unset arrow

set key bottom left reverse Left samplen 2
set lmargin screen 0.64
set rmargin screen 0.98
set xrange [0:1]
set yrange [0:1.01]
set xtics autofreq
set ytics autofreq
set format y '\tiny{set ylabel '\small{$r_\text{tr}$}' rotate by 0 offset 4
set xlabel '\small{$p$-quantile}'
#plot 'data/tcp-ex-coinc.dat' u ($0/4096):($4/32) w l lc '#333333' lw 6 dt 2 title '\small{expected}'
plot 'data/tcp-ex-coinc.dat' u ($0/4096):($2/32) w l lc 'purple'  lw 3 title '\small{independent}', \
     'data/tcp-ex-coinc.dat' u ($0/4096):($3/32) w l lc '#800080'    lw 3 title '\small{dependent}'
\end{gnuplot}
\end{center}
\caption{Canonical trigger coincidence processes (left) and corresponding QTR plot (right),
for a simulated time series paired with two event series
(independent and dependent, excerpts shown on top).
We generated the time series of length $T=4096$ from iid exponential random variables,
applied a moving average (MA) filter of order 8, standardized and subtracted the
minimum to obtain a non-negative time series $\mathcal{X}$ with serial dependencies.
We then generated two event series: an independent and a dependent one.
To simulate a peak trigger relationship, we randomly sampled $N_\mathcal{E} = 32$ time steps $t$
from the time series where $X_t > 4$, and set $E_{t-4}=1$ for these $t$ in the dependent
event series.
In the independent event series, we distributed the 32 events completely at random.
The time tolerance is set to $\Delta=7$.
}
\label{figTCP}
\end{figure}

Let $\bm{\tau} = (\tau_1, ..., \tau_M)$ be a sequence of increasing thresholds
$\tau_1 < ... <\tau_M$. The \textbf{trigger coincidence process}
\begin{equation}
\mathcal{K}_\text{tr} = \mathcal{K}^{\Delta,\bm{\tau}}_\text{tr}(\mathcal{E},\mathcal{X})
= \left(K^{\Delta,\tau_1}_\text{tr}, ..., K^{\Delta,\tau_M}_\text{tr}\right)
\end{equation}
is the corresponding sequence of the numbers of trigger coincidences for all
given thresholds~$\bm{\tau}$. A trigger coincidence process is always monotonically decreasing.
The \textbf{canonical trigger coincidence process} is given by the specific threshold sequence
$\bm{\tau} = (\tau_1,...,\tau_M) = (X_{(1)},...,X_{(T)})$,
where $X_{(t)}$ denotes the order statistic of the time series such that
$X_{(1)} < ... < X_{(T)}$. Trigger coincidence processes for other sequences of thresholds
approximate the canonical trigger coincidence process.
Figure~\ref{figTCP} (bottom left) illustrates the concept in a simulated example; simulation details
are in the caption.
At low thresholds, large numbers of trigger coincidences are observed both for the dependent
and the independent event series.
For higher thresholds, the numbers of trigger coincidences for the dependent event series
dramatically exceed the numbers of the independent event series.
By construction, all 32 events in the dependent series trigger an exceedance of the
threshold~4; see the marker ($\ast$).
The threshold~5 is exceeded after 13 out of 32 events from the dependent event series,
while it is only exceeded after a single event from the independent event series;
see the marker $(\dagger)$.

\subsubsection{Quantile-trigger rate plots}

Plots of trigger coincidence processes as in Figure~\ref{figTCP} (bottom left) help
to visually assess whether events in $\mathcal{E}$ systematically trigger
peaks of various magnitudes in a time series $\mathcal{X}$.
However, the scales of the axes depend on the range of values in $\mathcal{X}$
and the number of events $N_\mathcal{E}$, which makes it hard to compare these
plots across multiple pairs of time series and event series.
Furthermore, the absolute threshold value is not informative about the
actual extremeness of a peak with respect to the bulk of the data.
Therefore, we propose \textbf{quantile-trigger rate (QTR) plots}
as a standardized visualization of trigger coincidence processes with normalized axes.
In a QTR plot, the $x$-axis is normalized by using empirical $p$-quantiles from
$\mathcal{X}$ instead of the absolute thresholds $\tau_m$, while the $y$-axis
is normalized by using the trigger coincidence rate $r_\text{tr}$ instead
of the absolute number of trigger coincidences $K_\text{tr}$. The QTR plot
for the example above is shown in Figure~\ref{figTCP} (bottom right).
The most striking difference is that now the dependent curve appears more
extreme, since the thresholds larger than 4 correspond to high empirical
$p$-quantiles.
Intuitively, the closer an observed trigger coincidence
process to the top-right corner of the QTR plot, the more events coincide with
threshold exceedances, at more extreme levels.

\begin{figure*}[tbp]
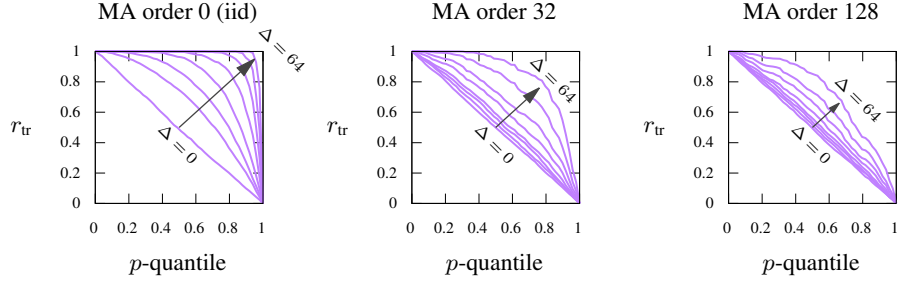

\begin{center}
\begin{gnuplot}[terminal=cairolatex,terminaloptions={size 4.9823,1.6 dashed}]
set multiplot layout 1,3
set xrange [0:1]
set yrange [0:1]
set ytics 0,0.2
set xlabel '$p$-quantile'
set ylabel '' rotate by 0
set lmargin 8
set format xy '\tiny{

set title 'MA order 0 (iid)'
set arrow from 0.5, 0.5 to 0.95, 0.95 lw 2 lc rgb '#444444' head filled front
set label '\tiny{$\Delta=0$}' at 0.4, 0.5 front rotate by -45
set label '\tiny{$\Delta=64$}' at 1.0, 1.15 front rotate by -45
set label '$r_\text{tr}$' at -0.5,0.5
plot "data/qtr-order-delta-new.dat" u 1:2 w l lw 3 lc 'purple' notitle, \
     "data/qtr-order-delta-new.dat" u 1:3 w l lw 3 lc 'purple' notitle, \
     "data/qtr-order-delta-new.dat" u 1:4 w l lw 3 lc 'purple' notitle, \
     "data/qtr-order-delta-new.dat" u 1:5 w l lw 3 lc 'purple' notitle, \
     "data/qtr-order-delta-new.dat" u 1:6 w l lw 3 lc 'purple' notitle, \
     "data/qtr-order-delta-new.dat" u 1:7 w l lw 3 lc 'purple' notitle, \
     "data/qtr-order-delta-new.dat" u 1:8 w l lw 3 lc 'purple' notitle, \
     "data/qtr-order-delta-new.dat" u 1:9 w l lw 3 lc 'purple' notitle
unset arrow
unset label

set title 'MA order 32'
set arrow from 0.5, 0.5 to 0.76, 0.76 lw 2 lc rgb '#444444' head filled front
set label '\tiny{$\Delta=0$}' at 0.4, 0.5 front rotate by -45
set label '\tiny{$\Delta=64$}' at 0.73, 0.95 front rotate by -45
set label '$r_\text{tr}$' at -0.5,0.5
plot "data/qtr-order-delta-new.dat" u 1:10 w l lw 3 lc 'purple' notitle, \
     "data/qtr-order-delta-new.dat" u 1:11 w l lw 3 lc 'purple' notitle, \
     "data/qtr-order-delta-new.dat" u 1:12 w l lw 3 lc 'purple' notitle, \
     "data/qtr-order-delta-new.dat" u 1:13 w l lw 3 lc 'purple' notitle, \
     "data/qtr-order-delta-new.dat" u 1:14 w l lw 3 lc 'purple' notitle, \
     "data/qtr-order-delta-new.dat" u 1:15 w l lw 3 lc 'purple' notitle, \
     "data/qtr-order-delta-new.dat" u 1:16 w l lw 3 lc 'purple' notitle, \
     "data/qtr-order-delta-new.dat" u 1:17 w l lw 3 lc 'purple' notitle
unset arrow
unset label

set title 'MA order 128'
set arrow from 0.5, 0.5 to 0.66, 0.66 lw 2 lc rgb '#444444' head filled front
set label '\tiny{$\Delta=0$}' at 0.4, 0.5 front rotate by -45
set label '\tiny{$\Delta=64$}' at 0.66, 0.82 front rotate by -45
set label '$r_\text{tr}$' at -0.5,0.5
plot "data/qtr-order-delta-new.dat" u 1:18 w l lw 3 lc 'purple' notitle, \
     "data/qtr-order-delta-new.dat" u 1:19 w l lw 3 lc 'purple' notitle, \
     "data/qtr-order-delta-new.dat" u 1:20 w l lw 3 lc 'purple' notitle, \
     "data/qtr-order-delta-new.dat" u 1:21 w l lw 3 lc 'purple' notitle, \
     "data/qtr-order-delta-new.dat" u 1:22 w l lw 3 lc 'purple' notitle, \
     "data/qtr-order-delta-new.dat" u 1:23 w l lw 3 lc 'purple' notitle, \
     "data/qtr-order-delta-new.dat" u 1:24 w l lw 3 lc 'purple' notitle, \
     "data/qtr-order-delta-new.dat" u 1:25 w l lw 3 lc 'purple' notitle
\end{gnuplot}
\caption{
Expected QTR plots for three time series with different levels of
serial dependencies (MA orders 0, 32, 128) and independent event series.
For every MA order, we simulate a single time series of length $T=4096$
from the exponential moving average model described above, and select
50 thresholds at equally spaced $p$-quantiles between 0 and 1.
For every threshold $\tau$ and every $\Delta \in \{0,1,2,4,8,16,32,64\}$,
we estimate the expected trigger coincidence rate
$r_\text{tr}^{\Delta,\tau} = K_\text{tr}^{\Delta,\tau}/N_\mathcal{E}$
for an independent event series by simulating 100~independent event series with
$N_\mathcal{E}=32$ events and averaging the trigger coicidence rates over the 100 runs.
Note that for large $\Delta$ and $\tau$, this expectation
can be approximated by the expected value of our GEV-based binomial distribution
from Equation~\ref{eqnKtrGEV}.
}
\label{figQTRplotsSimul}
\end{center}
\end{figure*}

However, QTR plots have to be interpreted with care.
The shape of a trigger coincidence process for an \textit{independent} pair of
event series and time series in a QTR plot depends on the statistical properties
of the input data.
For example, if $\mathcal{X}$ is an iid time series and $\mathcal{E}$ an iid Bernoulli
process, the fraction of events that coincide with an exceedance of the empirical
$p$-quantile of $\mathcal{X}$ (with time tolerance $\Delta=0$) is exactly $1-p$, and
the trigger coincidence process is a straight line from $(0,1)$ to $(1,0)$ in the QTR plot.
Figure~\ref{figQTRplotsSimul} illustrates the impact of serial dependencies in $\mathcal{X}$
and increasing time tolerance $\Delta$ on the shape of the trigger coincidence process under
independence in a QTR plot.
With increasing time tolerance $\Delta$, there are more trigger coincidences
under independence, and the lines in the QTR plot move towards the top-right corner
of the plot. This effect is strongest for iid time series, but also occurs for time
series with serial dependencies. Thus, a line that bends towards the top-right corner of
the QTR plot is necessary, but not sufficient to conclude a trigger relationship.
We need a statistical test that operates on the trigger coincidence process to assess
whether the shape in a QTR plot is unusual under an independence assumption.
For this purpose, we now derive the statistical properties of the trigger coincidence process.

\subsubsection{Markov model}
\label{secMarkovChainModel}
To assess whether a trigger coincidence process is so unusual that the null hypothesis of
independence has to be rejected, we consider the joint distribution
$\mathbb{P}(\mathcal{K}_\text{tr})
    = \mathbb{P}(K^{\Delta,\tau_1}_\text{tr}, ..., K^{\Delta,\tau_M}_\text{tr})$.
The product rule yields
\begin{gather}
\mathbb{P}(\mathcal{K}_\text{tr}) =
\mathbb{P}\left(K^{\Delta,\tau_1}_\text{tr}\right) \cdot \prod_{i=2}^{M} \mathbb{P}\left(K^{\Delta,\tau_i}_\text{tr}
             \mid K^{\Delta,\tau_1}_\text{tr}, ..., K^{\Delta,\tau_{i-1}}_\text{tr} \right).
\label{eqnTCPProductRule}
\end{gather}
We have already derived the marginal distribution $\mathbb{P}(K^{\Delta,\tau}_\text{tr})$
in Equation~\ref{eqnKtrGEV} and now focus on the conditionals.
Suppose there is an exceedance of $\tau_{i-1}$ in $\mathcal{X}$ within the window
$t,...,t+\Delta$, i.e., $Z_t^{\Delta,\tau_{i-1}}=1$.
The probability that there is also an exceedance of the higher threshold $\tau_i$ is
\begin{align}
\mathbb{P}(Z_t^{\Delta,\tau_{i}}=1 \mid Z_t^{\Delta,\tau_{i-1}}=1)
  = \frac{\mathbb{P}(Z_t^{\Delta,\tau_{i}}=1)}{\mathbb{P}(Z_t^{\Delta,\tau_{i-1}}=1)}
\approx \frac{1 - G(\tau_i ; \bm{\theta}_{\Delta})}{1 - G(\tau_{i-1} ; \bm{\theta}_{\Delta})},
\label{eqnMarkovCondSuccess}
\end{align}
where we used Equation~\ref{eqnBlockExceedanceGEV} for the approximation.
Equation~\ref{eqnMarkovCondSuccess} is valid whenever the
marginal approximation by the GEV is admissible.
We now rewrite the conditional random variable
$K^{\Delta,\tau_{i-1}}_\text{tr} \mid K^{\Delta,\tau_{i}}_\text{tr}$ by restricting the summation
from Equation~\ref{eqnTriggerCoincidencesMaxSimple} to time steps with both $E_t=1$ and
$Z^{\Delta,\tau_{i-1}}_t=1$ to incorporate our additional knowledge:
\begin{align}
K^{\Delta,\tau_{i}}_\text{tr} \mid K^{\Delta,\tau_{i-1}}_\text{tr} = \sum_{t:E_t=1, Z^{\Delta,\tau_{i-1}}_t=1} Z^{\Delta,\tau_i}_t.
\end{align}
As in Section~\ref{secTriggerDistrFixedThresh} we have a sum of identically distributed
Bernoulli trials, with success probability now given by Equation~\ref{eqnMarkovCondSuccess}
due to the additional knowledge.
Under the absence of long-range dependencies, the individual variables $Z^{\Delta,\tau_i}_t$
are approximately independent, and the conditional number of trigger
coincidences follows the binomial distribution
\begin{equation}
\label{eqnExceedanceProcessTransitionBinom}
K^{\Delta,\tau_i}_\text{tr} \mid K^{\Delta,\tau_{i-1}}_\text{tr} \sim \operatorname{Binomial}\left(K^{\Delta,\tau_{i-1}}_\text{tr},
    \frac{1 - \operatorname{G}(\tau_i ; \bm{\theta}_{\Delta})}{1 - \operatorname{G}(\tau_{i-1} ; \bm{\theta}_{\Delta})}\right).
\end{equation}
We thus rewrite the conditional distributions in Equation~\ref{eqnTCPProductRule}
to a first-order Markov structure
$\mathbb{P}(K^{\Delta,\tau_i}_\text{tr} \mid K^{\Delta,\tau_1}_\text{tr}, ..., K^{\Delta,\tau_{i-1}}_\text{tr})
  = \mathbb{P}(K^{\Delta,\tau_i}_\text{tr} \mid K^{\Delta,\tau_{i-1}}_\text{tr})$.
The joint probability $\mathbb{P}(\mathcal{K}_\text{tr})$ of the trigger coincidence process
under the null hypothesis of independence is then fully described by
Equation~\ref{eqnKtrGEV} for the smallest threshold and
Equation~\ref{eqnExceedanceProcessTransitionBinom} for all larger thresholds.

\subsubsection{Statistical test procedures}
With the results from Sections~\ref{secTriggerDistrFixedThresh} and~\ref{secMarkovChainModel},
we can now devise two test procedures to collect empirical evidence against
the null hypothesis of independence, in favor of the alternative hypothesis that events
systematically trigger exceedances at various thresholds.
\begin{enumerate}
\item We can employ our test procedure for \emph{pointwise} exceedances of individual thresholds
\emph{multiple times} at all given thresholds and adjust the resulting $p$-values using
methods for multiple hypothesis testing \citep{Dudoit2007}.
A potential shortcoming of this procedure is that the dependency
structure of trigger coincidence processes is ignored.
\item We can compute the joint probability of the observed trigger coincidence process and
reject the null hypothesis if the whole process is unusually unlikely under
the null distribution, in the sense specified below.
This approach takes the full dependency structure into account, but requires
Monte Carlo simulations. We refer to it as the \emph{multiple threshold test}.
\end{enumerate}
For the second approach, observe that the trigger coincidence process is a high-dimensional
discrete random variable, where every single realization---even the mode of the
distribution---has a very small likelihood.
We have to assess whether the observed likelihood
is \emph{unusually small} with respect
to the \emph{distribution of the likelihood values} under independence, i.e.,
we treat the likelihood as a random variable.
For numerical reasons, we work with the negative log-likelihood
$S(\mathcal{K}_\text{tr}) = -\log \mathbb{P}(\mathcal{K}_\text{tr})$
instead of the likelihood.
Formally, we use $S$ as a test statistic and reject the null hypothesis of independence
at significance level $\alpha$
if the $p$-value $\mathbb{P}(S \ge s) < \alpha$, where
$s = S(\mathcal{K}_\text{tr}(\mathcal{E},\mathcal{X}))$ is the observed value.
We use Monte Carlo simulations to approximate this $p$-value.
For this purpose, we generate $R$ independent event series $\mathcal{E}'$ with the same
number of events as $\mathcal{E}$ by randomly permuting $\mathcal{E}$.
For each independent event series, we determine the test statistic value
$s' = S(\mathcal{K}_\text{tr}(\mathcal{E'},\mathcal{X}))$ and compute the
Monte Carlo $p$-value \citep{Davison1997} via
$\hat{p} = \frac{1+|\{s' | s' \ge s\}|}{R+1}$.

 \section{Results and discussion}
\label{secResults}
We now utilize our statistical methodology to analyze whether severe Islamist terrorist attacks in
Western Europe and North America systematically trigger bursts of hate spech or counter-hate speech
on Twitter.
Additional simulations that support the methodological contributions described above can be found
in Appendix~\ref{secSimul}.

\subsection{Setup}
We use the data described in Section~\ref{secData}, which spans a total of $T=1,096$ days with
$N_\mathcal{E} = 17$ events.
We choose a time tolerance of $\Delta=7$ days to allow enough time for the news about
the incidents to spread globally. The simulation study in
Section~\ref{secNullDistributionPerformance} shows that this time tolerance is also large enough
for our GEV-based null distributions to be accurate.
For every social media time series $\mathcal{X}_i$, we estimate a GEV distribution $G_i$ by
splitting $\mathcal{X}_i$ into consecutive blocks of size $\Delta+1$ and fitting the parameters
of $G_i$ to the block maxima by maximum likelihood estimation.
We then select $M=32$ thresholds $\bm{\tau}_i = (\tau_{i,1},...,\tau_{i,32})$ at equidistant
$p$-quantiles between .75 and 1 from $\mathcal{X}_i$, and use the GEV distribution $G_i$ to obtain
the parameters of the binomial distributions
described by Equations~\ref{eqnKtrGEV} and~\ref{eqnExceedanceProcessTransitionBinom}.
We compute the observed trigger coincidence processes between the terrorist attack
event series $\mathcal{E}$ and all social media time series $\mathcal{X}_i$, and obtain
the respective test statistic values $s_i$ for the multiple threshold test.
We compute Monte Carlo $p$-values with $R=10{,}000$ simulations.
QTR plots with the Monte Carlo $p$-values are depicted in
Figure~\ref{figQTRplots}. The plots are augmented with the marginally
expected trigger coincidence processes under independence and the marginal 95\% confidence
intervals to additionally assess \textit{pointwise} exceedances of individual thresholds.

\subsection{Results}
The analysis shows that Islamist terrorist attacks in Western Europe
and North America \emph{systematically} trigger bursts of anti-Muslim hate speech on Twitter
(\texttt{\#stopislam}, $\hat{p}=.0317$).
90\% of Islamist terrorist attacks triggered an exceedance of the .85-quantile,
and 60\% of Islamist terrorist attacks even triggered an exceedance of the .95-quantile.
Our results confirm the findings of previous quantitative studies
\cite{Burnap2014,Magdy2015,Olteanu2018} with
a novel statistical methodology and a larger study period:
\textit{there is a clear systematic relationship between Islamist extremist violence offline
and anti-Muslim hate speech online.}

\begin{figure}[tbp]
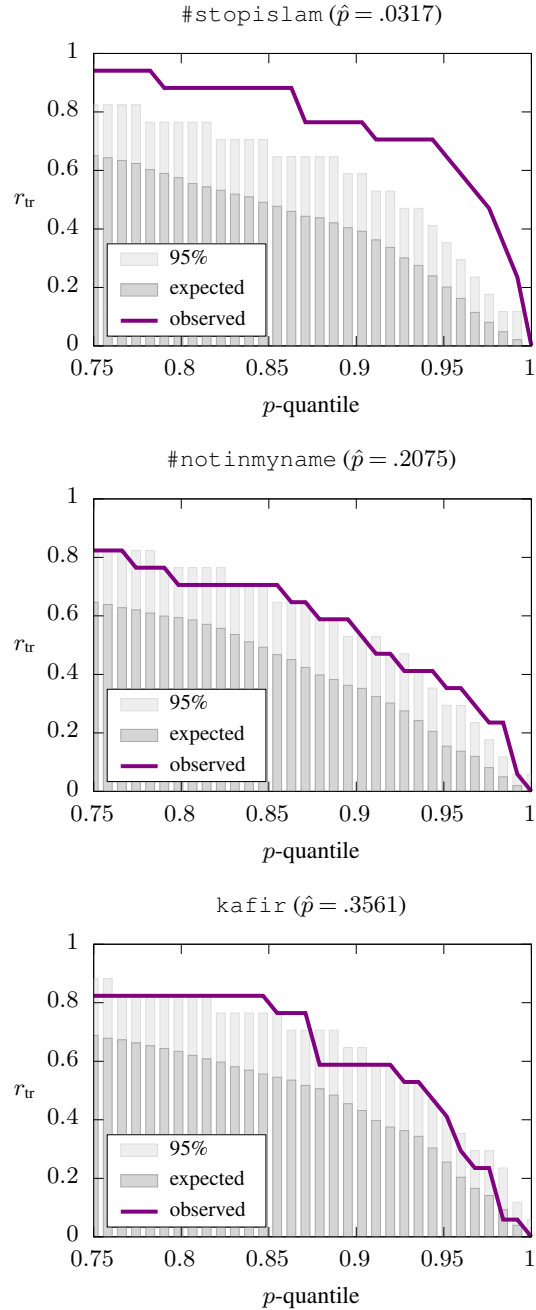

\begin{center}
\begin{gnuplot}[terminal=cairolatex,terminaloptions={size 3.,7 dashed}]
set multiplot layout 3,1
set xrange [0.75:1]
set yrange [0:1]
set xlabel '$p$-quantile'
set ylabel '$r_\text{tr}$' rotate by 0 offset 3
set lmargin 8
set rmargin 2
set grid front
unset grid

set boxwidth 0.005
set style fill solid 0.5

set key bottom left samplen 2 reverse Left box opaque spacing 1.25 width -2

set title '\texttt{\#stopislam} ($\hat{p}=.0317$)'
plot "data/eca-#stopislam-new.dat" u 1:($4/17) w boxes lc rgb '#DDDDDD' title '\scriptsize{95\% }', \
     "data/eca-#stopislam-new.dat" u 1:($3/17) w boxes lc rgb '#AAAAAA' title '\scriptsize{expected}', \
     "data/eca-#stopislam-new.dat" u 1:($2/17) w l lw 6 lc '#800080' title '\scriptsize{observed}'

set title '\texttt{\#notinmyname} ($\hat{p}=.2075$)'
plot "data/eca-#notinmyname-new.dat" u 1:($4/17) w boxes lc rgb '#DDDDDD' title '\scriptsize{95\%}', \
     "data/eca-#notinmyname-new.dat" u 1:($3/17) w boxes lc rgb '#AAAAAA' title '\scriptsize{expected}', \
     "data/eca-#notinmyname-new.dat" u 1:($2/17) w l lw 6 lc '#800080' title '\scriptsize{observed}'

set title '\texttt{kafir} ($\hat{p}=.3561$)'
plot "data/eca-kafir-new.dat" u 1:($4/17) w boxes lc rgb '#DDDDDD' title '\scriptsize{95\%}', \
     "data/eca-kafir-new.dat" u 1:($3/17) w boxes lc rgb '#AAAAAA' title '\scriptsize{expected}', \
     "data/eca-kafir-new.dat" u 1:($2/17) w l lw 6 lc '#800080' title '\scriptsize{observed}'
#$
\end{gnuplot}
\caption{QTR plots for severe Islamist terrorist attacks and their online social media response.}
\label{figQTRplots}
\end{center}
\end{figure}

On the other hand, our analysis \emph{does not} provide evidence for a \emph{systematic} association
between Islamist terrorist attacks and peaks in
jihadist hate speech (\texttt{kafir}, $\hat{p}=.2075$) or
Muslim counter-hate speech (\texttt{\#notinmyname}, $\hat{p}=.3561$) in the study period.
We stress that individual terrorist attacks may still have triggered such a social media
response.
Visual inspection of the data in Figure~\ref{figData} suggests peaks in the
hashtag \texttt{\#notinmyname} for Islamist terrorist attacks before July 2016.
Hashtag usage is typically subject to trends, so a systematic relationship can only be established
for hashtags that are used consistently throughout the study period.
The impact of an \emph{individual} terrorist attack on the social media time series can be assessed,
e.g., with counterfactual analyses \citep{Brodersen2015}, regardless of trends.

Figure~\ref{figQTRplots} also shows that even for jihadist hate speech and Muslim counter-hate
speech, the observed numbers of trigger coincidences fall outside the \emph{pointwise} 95\%
confidence intervals for some thresholds.
Pointwise tests at these specific thresholds would reject the null hypothesis of independence
on the basis of only a narrow perspective on the total association.
The multiple threshold test thus decreases the dangers of data dredging at the cost
of a lower sensitivity.
To validate the results from the multiple threshold test, we computed all $p$-values for
the pointwise tests at all thresholds and used different adjustment methods that control the
family-wise error rate at level $\alpha=.05$: Bonferroni,
single-step \v{S}id\'ak, step-down Holm in its original variant
and in the \v{S}id\'ak variant \citep{Dudoit2007}.
For all multiple test adjustment methods, the results agree with our multiple threshold test.

\subsection{Sensitivity analysis}
To assess the stability of the results, we further experimented with different choices for the
time tolerance $\Delta=4...16$ (ceteris paribus). We found that for $\Delta=4...8$,
the results of all tests on all time series are unchanged.
For $\Delta=9...14$ our multiple threshold test fails to reject the null hypothesis for the
\texttt{\#stopislam} time series, while the multiple pointwise test procedures still reject.
For $\Delta=15$ only the \v{S}id\'ak procedures reject the null hypothesis on
\texttt{\#stopislam}, while for $\Delta=16$ no test procedure rejects the null hypotheses
on any time series.
Choosing a time tolerance $\Delta$ that is longer than necessary thus reduces the
sensitivity of the tests.
We also varied the number of thresholds $M$ between 8 and 64 (ceteris paribus), which did
not change the outcome of any test. At last, we moved the thresholds upwards to more extreme levels
by choosing equidistant $p$-quantiles from the ranges .85 to 1 and .95 to 1, respectively
(ceteris paribus).
The outcomes on the \texttt{\#stopislam} and \texttt{kafir} time series remain unchanged, while
our multiple threshold test now detects an additional trigger relationship for
\texttt{\#notinmyname} that is not detected by the multiple pointwise test procedures.
Overall, the trigger relationship for \texttt{\#stopislam} is very stable across all test
procedures with different parameterizations, whereas the results on \texttt{\#notinmyname}
are inconclusive.

 \section{Conclusions}

We have refined the statistical methodology to infer potential causal links between
an event series and peaks in a time series. Based on the framework of event coincidence analysis,
the tests focus only the \textit{timing} of events and peaks, and no other distributional
characteristics.
We have derived analytical expressions for the null distributions of the ECA
statistic for coincidences with exceedances of a single threshold and multiple thresholds.
Our results are valid if the lagging time series satisfies the regularity conditions of the
Extremal Types Theorem.
We further require event occurrences in the leading event series to be separated by a sufficient
number of time steps such that the binomial null distributions are valid.
Our analysis is therefore most suitable for sparse event series.
For a complete causal analysis, confounding factors that influence both series
must still be ruled out. This is a challenging direction for future work, as it either requires
the specification of a joint model for the confounding factors, the time series, and the
event series in the spirit of Granger causality \citep{Granger1969}, or non-trivial changes
in the nonparametric procedure of ECA. A first step in this direction is conditional ECA and
joint ECA \citep{Siegmund2016a}.

\begin{appendix}
\section{Description of social media data}
\label{secDetData}

The hashtag \texttt{\#stopislam} has been observed in anti-Muslim hate
speech before \citep{Magdy2015,Olteanu2018} and has also received some media attention
\citep{Dewey2016,Hemmings2016}. Many posts that contain the hashtag actually condemn its usage,
so spikes in the volume should not be seen as pure bursts of hate speech. Yet, such condemnation
is typically triggered by initial anti-Muslim posts. Due to the mixed usage, the
magnitude of a spike is no indicator for the \textit{extent} of online hate, only the
\textit{presence} of a spike is informative.

The phrase ``not in my name'' is used by members of a group
to express their disapproval of actions that are associated with that group or
(perceived or actual) representatives of the group \citep{Tormey2006,Cehajic2008}. It was
observed, for example, during global protests against the 2003 war of the US-led coalition against
Iraq \citep{Bennett2004a}, or more recently during protests sparked by the murder of a Muslim boy
by Hindu nationalists in India 2017 \citep{Krishnan2017}.
Most importantly for the present study, Muslim social media users have repeatedly used the hashtag
after Islamist terrorist attacks \citep{Davidson2014}.
Due to the generic nature of the phrase, it cannot solely be viewed
as Muslim counter-hate speech. Nonetheless, online social media posts that contain
\texttt{\#notinmyname}
right after Islamist terrorist attacks are likely to convey a Muslim counter-hate message.

At last, the Arabic word \texttt{kafir} translates to the English word ``non-believer.''
It is traditionally used by Muslim fundamentalists against other Muslims that do not adhere to
the fundamentalist ideology \citep{Alvi2014}, but also against non-Muslims \citep{Bartlett2012},
in both cases to justify their killing.
The occurrence of the keyword \texttt{kafir} within online social media posts was recently shown
to be a strong indicator for jihadist hate speech \citep{DeSmedt2018}.
We use male, female and plural forms (\texttt{kafir}---\texttt{kafirah}---\texttt{kuffar}) in
Arabic script for the query.

We used the ForSight platform provided by Crimson
Hexagon\footnote{\url{https://www.crimsonhexagon.com/}}
to retrieve daily time series of the global Twitter volume for our keywords.
We excluded posts with the keyword \texttt{RT} to ignore retweets.
The time series are based on the full Twitter stream, which makes the numbers exact.
We preprocessed the original time series by taking the logarithm to base 2 and
subtracting the running mean over the past 30 days to make them stationary.

 \section{Simulation study}
\label{secSimul}

\subsection{Comparison of the null distributions}
\label{secNullDistributionPerformance}
The central result from Section~\ref{secTriggerDistrFixedThresh} is that under the
null hypothesis of independence (and some constraints on the time series), the number
of trigger coincidences for a single threshold approximately follows the binomial distribution
from Equation~\ref{eqnKtrGEV}, where the success probability is obtained from a GEV distribution.
This approximate result is useful specifically for the case of time series with serial
dependencies, where the Bernoulli-based null distribution from Equation~\ref{eqnDECANullDistrCond}
cannot be applied.
We now demonstrate that the Bernoulli-based null distribution indeed fails to describe the
empirically observed numbers of trigger coincidences for time series with serial dependencies,
while our GEV-based null distribution accurately describes the observed data.

\begin{figure}[tbp]
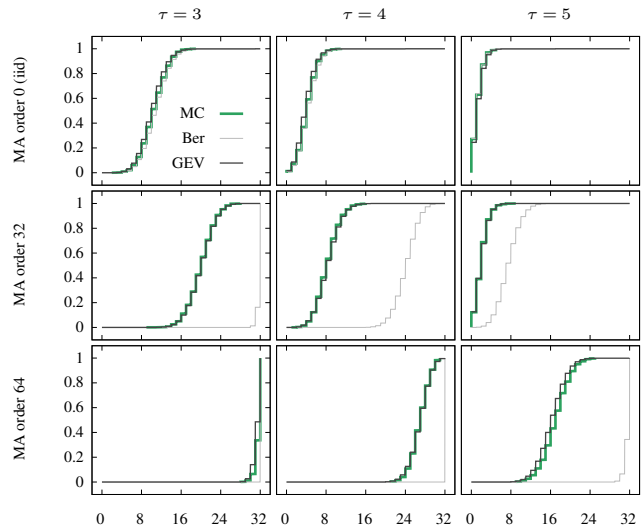

\begin{center}
\begin{gnuplot}[terminal=cairolatex,terminaloptions={size 3.3374,3}]
set multiplot layout 3,3

unset title
unset ylabel
unset xlabel
set xtics 0,8 scale 0.5
set border 15
set xrange [-2:34]
set yrange [-0.1:1.1]
set format y '\tiny{set format x ''

set title '\tiny{$\tau=3$}' offset 0,-0.6
set ylabel '\tiny{MA order 0 (iid)}' offset 1.5,0
set ytics 0.0,0.2,1.0 offset 0.5,0 scale 0.5
set tmargin screen 0.96
set bmargin screen 0.70
set lmargin screen 0.12
set rmargin screen 0.40
set key bottom right samplen 1
plot 'data/simul-maX-ne32-delta7-tau3.0000-cmf.dat' u 1:4 w l lw 4 lc rgb '#2ca25f' title '\tiny{MC}', \
     'data/estim-maX-ne32-delta7-tau3.0000-cmf.dat' u 1:2 w l lw 1 lc rgb '#bbbbbb' title '\tiny{Ber}', \
     'data/estim-maX-ne32-delta7-tau3.0000-cmf.dat' u 1:3 w l lw 2 lc rgb '#444444' title '\tiny{GEV}'
unset ytics
unset ylabel
set title '\tiny{$\tau=4$}'
set lmargin screen 0.41
set rmargin screen 0.69
plot 'data/simul-maX-ne32-delta7-tau4.0000-cmf.dat' u 1:4 w l lw 4 lc rgb '#2ca25f' notitle, \
     'data/estim-maX-ne32-delta7-tau4.0000-cmf.dat' u 1:2 w l lw 1 lc rgb '#bbbbbb' notitle, \
     'data/estim-maX-ne32-delta7-tau4.0000-cmf.dat' u 1:3 w l lw 2 lc rgb '#444444' notitle
set title '\tiny{$\tau=5$}'
set lmargin screen 0.70
set rmargin screen 0.98
plot 'data/simul-maX-ne32-delta7-tau5.0000-cmf.dat' u 1:4 w l lw 4 lc rgb '#2ca25f' notitle, \
     'data/estim-maX-ne32-delta7-tau5.0000-cmf.dat' u 1:2 w l lw 1 lc rgb '#bbbbbb' notitle, \
     'data/estim-maX-ne32-delta7-tau5.0000-cmf.dat' u 1:3 w l lw 2 lc rgb '#444444' notitle
unset title

set ylabel '\tiny{MA order 32}' offset 1.5,0
set ytics 0,0.2 offset 0.5,0 scale 0.5
set tmargin screen 0.69
set bmargin screen 0.43
set lmargin screen 0.12
set rmargin screen 0.40
plot 'data/simul-maX-ne32-delta7-tau3.0000-cmf.dat' u 2:4 w l lw 4 dt 1 lc rgb '#2ca25f' notitle, \
     'data/estim-maX-ne32-delta7-tau3.0000-cmf.dat' u 1:4 w l lw 1 dt 1 lc rgb '#bbbbbb' notitle, \
     'data/estim-maX-ne32-delta7-tau3.0000-cmf.dat' u 1:5 w l lw 2 dt 1 lc rgb '#444444' notitle
unset ylabel
unset ytics
set lmargin screen 0.41
set rmargin screen 0.69
plot 'data/simul-maX-ne32-delta7-tau4.0000-cmf.dat' u 2:4 w l lw 4 dt 1 lc rgb '#2ca25f' notitle, \
     'data/estim-maX-ne32-delta7-tau4.0000-cmf.dat' u 1:4 w l lw 1 dt 1 lc rgb '#bbbbbb' notitle, \
     'data/estim-maX-ne32-delta7-tau4.0000-cmf.dat' u 1:5 w l lw 2 dt 1 lc rgb '#444444' notitle
set lmargin screen 0.70
set rmargin screen 0.98
plot 'data/simul-maX-ne32-delta7-tau5.0000-cmf.dat' u 2:4 w l lw 4 dt 1 lc rgb '#2ca25f' notitle, \
     'data/estim-maX-ne32-delta7-tau5.0000-cmf.dat' u 1:4 w l lw 1 dt 1 lc rgb '#bbbbbb' notitle, \
     'data/estim-maX-ne32-delta7-tau5.0000-cmf.dat' u 1:5 w l lw 2 dt 1 lc rgb '#444444' notitle

set ylabel '\tiny{MA order 64}' offset 1.5,0
set ytics 0,0.2 offset 0.5,0 scale 0.5
set xtics 0,8 scale 0.5
set format x '\tiny{set tmargin screen 0.42
set bmargin screen 0.16
set lmargin screen 0.12
set rmargin screen 0.40
plot 'data/simul-maX-ne32-delta7-tau3.0000-cmf.dat' u 3:4 w l lw 4 lc rgb '#2ca25f' notitle, \
     'data/estim-maX-ne32-delta7-tau3.0000-cmf.dat' u 1:6 w l lw 1 lc rgb '#bbbbbb' notitle, \
     'data/estim-maX-ne32-delta7-tau3.0000-cmf.dat' u 1:7 w l lw 2 lc rgb '#444444' notitle
unset ylabel
unset ytics
set lmargin screen 0.41
set rmargin screen 0.69
plot 'data/simul-maX-ne32-delta7-tau4.0000-cmf.dat' u 3:4 w l lw 4 lc rgb '#2ca25f' notitle, \
     'data/estim-maX-ne32-delta7-tau4.0000-cmf.dat' u 1:6 w l lw 1 lc rgb '#bbbbbb' notitle, \
     'data/estim-maX-ne32-delta7-tau4.0000-cmf.dat' u 1:7 w l lw 2 lc rgb '#444444' notitle
set lmargin screen 0.70
set rmargin screen 0.98
plot 'data/simul-maX-ne32-delta7-tau5.0000-cmf.dat' u 3:4 w l lw 4 lc rgb '#2ca25f' notitle, \
     'data/estim-maX-ne32-delta7-tau5.0000-cmf.dat' u 1:6 w l lw 1 lc rgb '#bbbbbb' notitle, \
     'data/estim-maX-ne32-delta7-tau5.0000-cmf.dat' u 1:7 w l lw 2 lc rgb '#444444' notitle
\end{gnuplot}
\caption{Cumulative probability mass functions for the number of trigger coincidences under
independence, obtained empirically by Monte Carlo simulations (MC), and analytically with the
Bernoulli-based binomial (Ber) and the GEV-based binomial (GEV).}
\label{figGEVperfMatrix}
\end{center}
\end{figure}

For this purpose, we simulate three time series with MA orders of 0, 32 and 64.
For every time series, we simulate 1,000 independent pairs of event series with
$N_\mathcal{E}=32$ events, and record the numbers of trigger coincidences at the
three thresholds $\tau \in \{3,4,5\}$, with time tolerance $\Delta=7$.
For every time series and choice of threshold, we compare the empirically obtained
(Monte Carlo) null distribution with the two analytical null distributions.
All cumulative probability mass functions are visualized in Figure~\ref{figGEVperfMatrix}.
The visualizations clearly show that our GEV-based estimate closely follows the empirical
distribution in all runs, while the Bernoulli-based estimate is only correct for iid time series.
These results also demonstrate that a value of $\Delta=7$ is already large enough for
the GEV approximation to be valid.

\subsection{Analysis of the Markov model}
\begin{figure}[tbp]
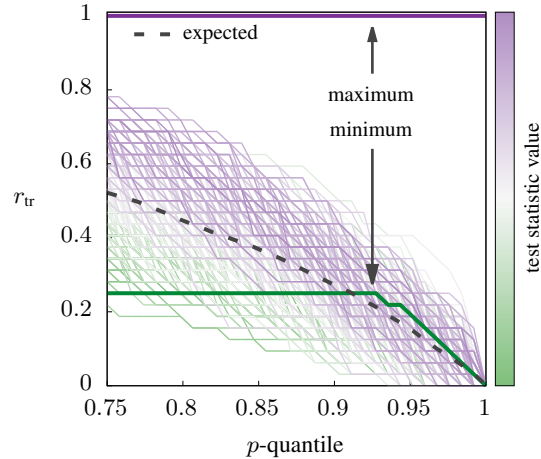

\begin{center}
\begin{gnuplot}[terminal=cairolatex,terminaloptions={size 3.3374,2.5}]
# read surprise values from file into array
FILE = 'data/simul-seqs-exp-O8-nlls-new.dat'
array Nll[1000]
stats FILE using (Nll[int($0+1)] = $1) prefix 'A'

set xrange [0.75:1]
set yrange [0:1.01]
set xlabel '$p$-quantile'
set ylabel '$r_\text{tr}$' rotate by 0 offset 2
set palette defined (0 '#99d594', 0.5 '#ffffbf', 1 '#fc8d59')
set palette defined (0 '#008837', 0.5 '#f7f7f7', 1 '#7b3294')
set palette defined (0 '#7fbf7b', 0.5 '#f7f7f7', 1 '#af8dc3')
set xtics scale 0.5 nomirror
set ytics scale 0.5 nomirror
set colorbox vertical
unset cbtics
set cblabel '\scriptsize{test statistic value}'
set key top left reverse Left samplen 2

set arrow from 0.925,(27./32) to 0.925,(31./32) lw 4 lc rgb '#444444' head filled front
set label '\scriptsize{maximum}' at 0.925,(25./32) front center
set arrow from 0.925,(20.5/32) to 0.925,0.275 lw 4 lc rgb '#444444' head filled front
set label '\scriptsize{minimum}' at 0.925,(22./32) front center
plot for [i=2:1000] 'data/simul-seqs-exp-O8-new.dat' u 1:(column(i)/32):(Nll[i-1]) w l lw 2 lc palette notitle, \
   'data/simul-seqs-exp-O8-minmaxsurp-new.dat' u 1:($2/32) w l lw 6 lc '#7b3294' notitle, \
   'data/simul-seqs-exp-O8-minmaxsurp-new.dat' u 1:($3/32) w l lw 6 lc '#008837' notitle, \
   'data/simul-seqs-exp-O8-minmaxsurp-new.dat' u 1:($4/32) w l lw 6 dt 2 lc '#444444' title '\scriptsize{expected}'
\end{gnuplot}
\caption{Simulated trigger coincidence processes under independence, colorized by the test
statistic value, along with the processes that attain the theoretical minimum and
maximum test statistic value.}
\label{figQTRplotsSurprise}
\end{center}
\end{figure}

Our second central result is the Markov model for trigger coincidence processes
from Section~\ref{secIncreasingThresholds}, with the associated test statistic
for the multiple threshold test.
Larger values of the test statistic should correspond with visually more ``extreme''
trigger coincidence processes in a QTR plot.
To confirm this assumption, we illustrate the test statistic values for a single simulated
time series (MA order 8) and 1,000 independent event series (with 32 events).
We use a time tolerance $\Delta=7$ and 32~thresholds at equally spaced $p$-quantiles
between .75 and 1 from the time series.
All resulting trigger coincidence processes are plotted in Figure~\ref{figQTRplotsSurprise},
colorized by their test statistic values.
We also plot the trigger coincidence process with the highest (lowest) test statistic value
that is theoretically possible; we obtain them by maximizing (minimizing) the test statistic
over all possible processes with a dynamic programming approach.
At last, we show the marginally expected trigger coincidence process at every threshold
$\tau_m$, i.e., the value of $\operatorname{E}[K^{\Delta,\tau_m}_\text{tr} \mid N_\mathcal{E}=32]$
obtained from Equation~\ref{eqnKtrGEV}.
All simulated trigger coincidence processes are close to the marginally expected sequence;
the more they bend towards the top-right corner of the plot, the higher the test statistic value.
The trigger coincidence process with the highest possible test statistic value closely traces
the top-right corner, which corresponds to our intuitive notion of the most unusual outcome:
all events trigger exceedances of the highest quantiles.

\end{appendix}

\bibliographystyle{imsart-nameyear}
\bibliography{library}

\end{document}